\documentclass{iaus}
\usepackage{graphicx}

\title[] %% give here short title %%
{Binary populations and stellar dynamics in young clusters}

\author[]  %% give here short author list %%
{D. Vanbeveren$^{1,2}$, H. Belkus$^{1}$, J. Van Bever$^3$ \and N. Mennekens$^{1}$}

\affiliation{$^1$Astrophysical Institute, Vrije Universiteit Brussel, Brussels, Belgium \\ email: {\tt dvbevere@vub.ac.be, hbelkus@vub.ac.be} \\[\affilskip]
$^2$GroupT, Association K.U. Leuven, Louvain, Belgium \\ email: {\tt dany.vanbeveren@groept.be}
\\[\affilskip]
$^3$Institute of Computational Astrophysics, St.-Mary's University, Halifax, Canada \\email: {\tt vanbever@penguin.stmarys.ca}}

\pubyear{2008}
\volume{xxx}  %% insert here IAU Symposium No.
\pagerange{119--126}
\jname{Title of your IAU Symposium}
\editors{A.C. Editor, B.D. Editor \& C.E. Editor, eds.}
\begin{document}

\maketitle

\begin{abstract}
We first summarize work that has been done on the effects of binaries on theoretical population synthesis of stars and stellar phenomena. Next, we highlight the influence of stellar dynamics in young clusters by discussing a few candidate UFOs (unconventionally formed objects) like intermediate mass black holes,  $\eta$ Car, $\zeta$ Pup, $\gamma$$^{2}$ Velorum and WR 140.


\end{abstract}

\firstsection % if your document starts with a section,
\section{Introduction}

\medskip

Many (most?) of the massive stars are formed in clusters and the observations reveal that many cluster members are binary components (to illustrate, see Sana et al., 2007). The effect of binaries on population studies has been discussed many times by different research groups (the Brussels group was one of them) but the following points may be worth repeating:

\medskip

\begin{itemize}
\item binaries with initial primary mass larger than 40 $M_{\odot}$ and with a period that is large enough so that LBV mass loss happens before the Roche lobe overflow (RLOF) would start, avoid RLOF (the LBV scenario as it was introduced in Vanbeveren, 1991); this may not be true for case A binaries
\item most of the primaries of binaries with orbital period as large as 10 yrs and with initial mass smaller than 40 $M_{\odot}$ lose most of their hydrogen rich layers by RLOF; the evolution of these stars is therefore quite different from the evolution of single stars with the same mass
\item the RLOF in late case B and case C binaries may be accompanied by the common envelope process that may result in the merger of the two components; it can be expected that this merger will be a rapid rotator with peculiar chemical abundances  
\item the RLOF in Case A and early case B binaries implies mass and angular momentum transfer towards the gainer; together with the merger process discussed above, this process is a natural way to make rapid rotators and therefore massive close binaries may be natural progenitors of long gamma ray bursts (Cantiello et al., 2007)
\item the favored model for short gamma ray bursts is the binary merger of two relativistic compact stars and it cannot be excluded that such mergers are important sites of galactic r-process enrichment (De Donder and Vanbeveren, 2003a) 
\item	the observed massive binary frequency is not necessarily the binary frequency at birth. This is due to the fact that a significant fraction of all massive binaries become single stars during their evolution (many binaries with small mass ratio merge, late case B and case C binaries evolve through a common envelope phase and may merge as well, the supernova explosion of one of the components disrupts the binary), e.g. many single stars may have a binary past with an evolution which is distinctly different from stars that are born as single stars.
\end{itemize} 

\medskip

The proceedings of the conference on 'Massive Stars in Interacting Binaries' (eds. A. Moffat and N. St.-Louis) were published a few months ago. They can be considered as an update of the monograph 'The Brightest Binaries' published in 1998 (Vanbeveren et al. 1998c) and we recommend careful reading to anyone with an open mind willing to admit that many theoretical population studies, where the effects of binaries are ignored, have mainly an academic value. In section 2 we list a few extra references that may be interesting in order to learn more about the effects of massive binaries on population studies.
 
Many massive objects (single or binary) are born in dense (embedded) clusters, containing a few 10s up to a few 1000s massive stars and even more (e.g., the clusters in the Carina association, the Arches and Quintuplet clusters in the galactic bulge, the Orion nebula cluster, R136 in the LMC, MGG-11 in M82, etc.). In these clusters the evolution of each object may be affected by the presence of all the others, or at least by the closest neighbors. In sections 3 and 4 we discuss a few peculiar massive objects that may be the result of the dynamical evolution of stars in dense clusters, we like to call them Unconventionally Formed Objects (UFOs).

\section{The effects of binaries on theoretical massive star population studies}
 
\medskip
 
\begin{itemize}
\item {\it The WR population}: in the sixties and seventies, before it was realized that LBV and/or RSG stellar wind mass loss rates are large enough to remove most of the hydrogen rich layers of a massive star, it was generally accepted that most of the WR stars are formed through binary mass loss processes. It may sound as a surprise, but the study of Vanbeveren and Conti (1980) was probably among the first where convincing arguments were presented that no more than 40-50\% of all galactic WR stars are binary members, indicating that massive single stars have to evolve into WR stars as well. This binary percentage still holds today. The effect of binaries on the WR population was described in detail in Vanbeveren et al. (1998a, b, c), Vanbeveren et al. (2007) (see also J. Eldridge in the present proceedings). Since all primaries of interacting binaries become hydrogen deficient stars at the beginning of core helium burning, all these primaries become WR-like stars. Therefore, a crutial parameter that affects theoretical WR-binary population studies is the minimum mass a binary component should have so that the WR-like star will be observed as a real WR star   
\item {\it The supernova rates}: the effect of binaries on supernova rates (types Ia and Ib/c) has been investigated in much detail by Tutukov et al. (1992), Belczynski et al. (2002), also in the framework of a galactic evolutionary model (De Donder and Vanbeveren, 2003b, 2004). More recent studies confirm this earlier work, although proper referencing is sometimes missing.
\item	{\it The Be stars}: many Be stars are the optical component of Be-X ray binaries, which may indicate that they became Be stars via the binary evolutionary channel (the binary mass transfer process). Accounting for the fact that the supernova explosion of a binary component disrupts most of the binaries, the large number of Be-X ray binaries may indicate that there are a much larger number of Be single stars which have a similar evolutionary past as the Be stars in the X-ray binaries. More information on how binaries affect the Be-star population is given in Pols et al. (1991), Pols and Marinus (1994), Van Bever and Vanbeveren (1997). In the latter paper it was argued that the rich population of Be stars in some clusters can not be explained by binary evolution alone.
\item	{\it Binaries and the spectral synthesis of massive starbursts}: Belkus et al. (2003) and Van Bever and Vanbeveren (2000, 2003) studied the effect of binaries on the spectral synthesis of massive starbursts. Note that binaries play a very important role when the starburst is at least 4-5 Myr old. 
\item	{\it Binaries and galactic chemical evolution}: many stars are formed in binaries and it therefore looks quite strange that most of the galactic chemical evolutionary studies account for single star chemical yields only. In Brussels we have a project where the effect of binaries on galactic evolution is studied in detail. Results have been published in an extended review by De Donder and Vanbeveren (2004) and in references quoted therein. Notice that as far as the C, N, O, ... Fe elements and r-process elements are concerned, binary yields are very different from single star yields and the effect on a galactic model is at least as large as the effect of fast rotation. But including binaries in a galactic code is not something as simple as replacing one set of yields by another.
\end{itemize}

\section{Very massive stars and intermediate mass black holes: UFOs of the first kind}

\medskip

The X-ray luminosity in ultra-luminous X-ray sources (ULXs) may be as high as 10$^{42}$ erg/s (Ptak and Colbert, 2004). Many (most) of these ULXs are found in young dense star clusters (e.g., the ULX in the cluster MGG-11 in M82), preferentially outside the cluster core. To understand these high luminosities by means of sub-Eddington mass accretion onto a relativistic object, one has to accept that a black hole (BH) component is present with a mass as high as 1000 $M_{\odot}$ (the term intermediate mass BH or IMBH is used). However, note that there exists an alternative model where the high luminosities are explained by  supra-Eddington accretion onto a BH with a much smaller mass of 50-100 $M_{\odot}$ (Soria, 2007 and references therein). 

The dynamical evolution of dense massive clusters (like MGG-11) obviously depends on the initial star density, but, most interestingly, is characterized by the processes of mass segregation and core collapse, which happen on a timescale similar to the evolutionary timescale of the most massive stars in the cluster. Core collapse is accompanied by real physical stellar collisions between the most massive stars and the possible formation of an intermediate mass object with a mass of 1000 $M_{\odot}$ or larger. It is therefore tempting to link the core collapse process and the existence of ULXs. This has been investigated in many papers by the team working with the direct N-body STARLAB software (e.g., Portegies Zwart et al, 2007, and references therein) and working with the statistical MONTE-CARLO software (e.g., Fregeau and Rasio, 2007, and references therein). 

The link with ULXs has two major uncertainties: core collapse is a fact and the successive collision of the most massive stars in the core (the term runaway merger is used) is a fact as well, but, when this merger process stops, it is at present unclear what happens with that merger object? Does it form a star with a mass similar to the mass of the merger object, and, when a very massive star is formed, how does it evolve? It is clear that when it forms, its evolution will be critically affected by stellar wind mass loss during core hydrogen burning (CHB) and during core helium burning (CHeB). 

Belkus et al. (2007) investigated the evolution of very massive stars with an initial mass up to 1000 Mo, with a metallicity Z between 0.001 and 0.04, using the CHB stellar wind mass loss rate formalism of Kudritzki (2002) and the CHeB wind formalism proposed by Vanbeveren et al. (1998a, b, c), which is very similar to the one proposed by Nugis and Lamers (2000). Belkus et al. concluded that very massive stars with solar or super solar metallicity end their life as a BH with a mass not larger than 70-80 $M_{\odot}$. Yungelson et al. (2008) recently published their study on the evolution of very massive stars with solar metallicity. Although they use 'ad hoc' (the term used by the authors) stellar wind mass loss rate formalisms, they essentially arrive at the same conclusion as Belkus et al. 

Belkus et al. presented an easy evolutionary recipe for very massive stars. The latter is combined with our own direct N-body code and with our massive star evolution handler to simulate the early evolution of MGG-11. We adopt a starburst model with 3000 massive stars with solar type metallicity and a King density distribution with a half mass radius of 0.5 pc (more details are given in Vanbeveren et al., 2008 and in Belkus et al., 2008). The main conclusion is that stellar wind mass loss during CHB does not prevent the formation of a very massive object, but when the merger process stops and when this object becomes a very massive star, stellar wind mass loss determines its further evolution and an IMBH is not formed. This argues against the sub-Eddington model of the ULX in MGG-11.

\section{UFOs of the second kind: why not $\eta$ Car, $\gamma$$^2$ Velorum, $\zeta$ Pup, WR 140?}

\medskip

When a (small or large) number of massive objects (single stars or binaries) form together in a cluster with small radius, the evolution of each member may be affected by the presence of the others. The dynamical evolution of such clusters implies close encounters leading to physical collisions and stellar mergers. The collision process of two massive stars (two 88 $M_{\odot}$ stars and a 88 $M_{\odot}$ star with a 28 $M_{\odot}$ star) has been investigated by Suzuki et al. (2007) by using SPH and the following conclusions are striking: 

\medskip

\begin{itemize}
\item the merging lasts a few days and during the merging process 10-12 $M_{\odot}$ are lost
\item the merger product is a mixed star where the degree of homogenization is larger in case both stars have similar masses. 
\end{itemize}

\medskip

\noindent {\it $\eta$ Car}: it is tempting to link the collision process discussed above and the loss of about 10 $M_{\odot}$ during the 19th century event of this object. If the $\eta$ Car progenitor experienced a dynamical encounter with a binary or with a single star that resulted in a stellar merger, the sudden very large mass loss and the very large eccentricity of the binary at present can be explained in a natural way.

\medskip

\noindent {\it $\gamma$$^2$ Velorum}: The formation of WR+OB binaries in young dense stellar systems may be quite different from the conventional binary evolutionary scenario. Mass segregation in dense clusters happens on a timescale of one or a few million years which is comparable to the evolutionary timescale of a massive star. Within the lifetime of a massive star, close encounters may therefore happen very frequently. When we observe a WR+OB binary in a dense cluster of stars, its progenitor evolution may be very hard to predict. Our simulations predict the following UFO-scenario of WR+OB binaries. After 4 million years the first WR stars are formed, either single or binary. Due to mass segregation, this happens most likely when the star is in the starburst core. Dynamical interaction with another massive object becomes probable, especially when the other object is a binary. We encountered a situation where a WR star (a single WC-type with a mass = 10 Mo) interacts with a 16 $M_{\odot}$ + 14 $M_{\odot}$ circularized binary with a period P = 6 days. We explored the result of such an encounter by using the FEWBODY software of Fregeau et al. (2004). The details of all these simulations will be discussed elsewhere (Belkus et al., 2008). One of the results is the following: the two binary components merge and the 30 $M_{\odot}$ merger forms a binary with the WC star with a period of 80 days and an eccentricity e = 0.3. This binary very well resembles the WR+OB binary  $\gamma$$^2$-Velorum but it is clear that conventional binary evolution has not played any role in its formation.

\medskip

\noindent {\it $\zeta$ Pup}: $\zeta$ Pup, $\lambda$ Cep and BD+43$^{o}$3654 are 3 massive runaways with a runaway velocity between 40 km/s and 70 km/s. Their location in the HR diagram suggests that they belong to the most massive star sample of the solar neighborhood (Vanbeveren et al., 1998b, c; Hoogerwerf et al., 2001; Comeron \& Pasquali, 2007). Runaways can be formed by the binary-SN scenario (Blaauw, 1961), where the original massive primary (the mass loser when the Roche lobe overflow process happens) explodes and eventually disrupts the binary, leaving a neutron star remnant and a runaway secondary (the mass gainer when the Roche lobe overflow happens). Such a scenario for $\zeta$ Pup was presented by Vanbeveren et al. (1998b, c). To explain the significant surface helium enrichment of the star, its rapid rotation and its runaway velocity (= 70 km/s), the mass transfer phase and the accretion process must be accompanied by spinning-up and quasi-homogenization of the mass gainer (the full mixing model as it was introduced by Vanbeveren and De Loore, 1994) whereas the overall evolution should have resulted in a pre-SN binary with a period of the order of 4 days. The latter requires some fine-tuning. 

To illustrate that the dynamical ejection mechanism is a very valuable alternative, the
FEWBODY software of Fregeau et al. (2004) was used to reproduce the observed properties of $\zeta$ Pup. We performed over 1 million single star-binary and binary-binary scattering experiments. The details of these experiments will also be  given in Belkus et al. (2008). We explored the effects of different masses and different binary periods and eccentricities and, obviously, many experiments reproduce $\zeta$ Pup, but to obtain  a runaway velocity as observed, the binaries participating in the scattering process always have to be very close (periods smaller than 100 days). Most interestingly, in all our experiments, {\it $\zeta$ Pup turns out to be a merger of 2 or 3 stars.}

\medskip

\noindent {\it WR 140}: this WC5 + O4-5 binary has a period of 7.9 yrs, an eccentricity e = 0.85 and minimum component masses = 23 + 62 $M_{\odot}$. The O4-5 star is much younger than the WC5 star. Accounting for the very large period of the binary and the very large masses of both components, a typical binary rejuvenation process during RLOF seems unlikely. The only alternative then is a dynamical process, e.g. single-binary or binary-binary. Interestingly, a dynamical process explains the eccentricity in a natural way and it is tempting to link WR 140 and  $\eta$ Car.

\section{An experiment}

\medskip

About 10\% of all the O-type stars in the solar neighborhood may be runaway stars with a peculiar space velocity $>$ 30-40 km/s (Gies and Bolton, 1986). Runaways can be formed by the binary-SN scenario (Blaauw, 1961) or by two-body dynamical interactions in star clusters where at least one body is a binary (Leonard and Duncan, 1990). It is clear that the number of O-type runaways predicted by both processes depends on the adopted binary frequency. We performed the following experiment:
 
\medskip

\noindent {\it calculate the O-type runaway frequency as function of the primordial binary cluster frequency accounting for cluster stellar dynamics and for the SN explosion in binaries, using typical initial parameters of solar neighborhood type clusters (like the Orion Nebula Cluster)}

\medskip

Preliminary results indicate that a 10\% O-type runaway frequency can be obtained only by a model where the primordial (interacting) massive binary frequency in the solar neighborhood is larger than 60\%.

\end{document}